\documentstyle[jkas]{article}

\beginpage{1}
\endpage{10}
\year{2013}\volume{00}\month{February}

\runningauthor {J.-Y. KIM \& S. TRIPPE}
\runningtitle{AGN INTRA-DAY VARIABILITY}

\month{February} \year{2013} \volume{00} \issueno{00}
\beginpage{1}\endpage{10}
\date{Received 2012 December 21; Revised 2013 February 5; Accepted 2013 February 5}

\begin{document}

%%% physics style sectioning %%%
%\def\thesection{\Roman{section}}
%\def\thesubsection{\Alph{subsection}.}
%\def\thesubsubsection{\arabic{subsubsection}.}

\title{HOW TO MONITOR AGN INTRA-DAY VARIABILITY AT 230\,GHZ}

\author{Jae-Young Kim \& Sascha Trippe\vspace{0.5mm}}

\address{Department of Physics and Astronomy, Seoul National University,
Seoul 151-742, South Korea \\{\it E-mail : dadu1543@snu.ac.kr; trippe@astro.snu.ac.kr}}

\address{\normalsize{\it Received 2012 December 21; Revised 2013 February 5; Accepted 2013 February 5}}

\abstract{\noindent We probe the feasibility of high-frequency radio observations of very rapid flux variations in compact active galactic nuclei (AGN). Our study assumes observations at 230\,GHz with a small 6-meter class observatory, using the SNU Radio Astronomical Observatory (SRAO) as example. We find that 33 radio-bright sources are observable with signal-to-noise ratios larger than ten.  We derive statistical detection limits via exhaustive Monte Carlo simulations assuming (a) periodic, and (b) episodic flaring flux variations on time-scales as small as tens of minutes. We conclude that a wide range of flux variations is observable. This makes high-frequency radio observations -- even with small observatories -- a powerful probe of AGN intra-day variability; especially, those observations complement observations at lower radio frequencies with larger observatories like the Korean VLBI Network (KVN).
}


\offprints{\sc S. Trippe}

\maketitle

\section{INTRODUCTION}

\noindent
Active galactic nuclei (AGN; for a review, see e.g. \citealt{beckmann2012}) show strong flux variations on various timescales. On the shortest timescales, intra-day variability (IDV) of radio fluxes was first reported by \citet{wit86}. Since then, such a rapid variability has been observed throughout the electromagnetic spectrum mainly in compact objects, such as flat spectrum radio quasars (FSRQs) and BL Lacertae objects (BL Lacs). At frequencies around 5\,GHz, about half of all FSRQs have been reported to show IDV \citep{lov08}. 

A priori, such a rapid variability is hard to understand. For a given observed variability time scale $t_{\rm var}$ the spatial extension of the emission region is limited to $c\,t_{\rm var} / (z + 1)$, with $c$ denoting the speed of light and $z$ being the redshift. The observed brightness temperature is related to the variability time scale like $T_b\propto\left[t_{\rm var}\,(z + 1)\right]^{-2}$; observed values are as high as $T_b\approx10^{18}$\,K \citep{wag95,fuhr08}. However, cooling by inverse Compton scattering limits the brightness temperature of realistic plasmas to $T_b\lesssim10^{12}$\,K \citep{kel69} -- in apparent contradiction to observations.

In order to understand intra-day variability, several intrinsic and extrinsic origins of IDV have been proposed. The most important \emph{extrinsic} mechanism is \emph{interstellar scintillation} (ISS): turbulent motion of ionized interstellar matter focuses and de-focuses the light from AGN on its way to Earth. ISS has been identified observationally by annual modulations of IDV caused by the motion of the Earth around the sun \citep{gab07,liu12,mar12}. However, the amplitude of interstellar scintillation scales with the observation frequency $\nu$ like $\nu^{-2.2}$ \citep{rickett1984}. Accordingly, the contribution by ISS to AGN variability is substantial at cm-radio frequencies but increasingly unimportant at mm-radio and shorter wavelengths. In addition, the observed correlation between rapid variability in radio and optical wavelength regimes indicates intrinsic causes \citep{wag95}.

The most important \emph{intrinsic} mechanism for intra-day variability is a combination of rapid intrinsic variability of the source -- especially shocks in AGN jets \citep{blan79} -- plus Doppler boosting. In order to reconcile the observed brightness temperatures with the inverse Compton limit, Doppler factors on the order of 10 to 100 are necessary -- which are high though still realistic values \citep{fuhr08}. 

An analysis of plasma-physical processes related to AGN outflows usually requires radio observations as the emission from AGN jets is -- essentially -- synchrotron radiation; the flux density $F_{\nu}$ approximately follows a power law $F_{\nu}\propto\nu^{-\alpha}$ with $\alpha\approx0...1$. Addressing intra-day variability requires flux monitoring observations that (1) operate at high frequencies $\nu\gtrsim100$\,GHz in order to bypass ISS, (2) have good ($\lesssim$1\,h) time resolution, (3) cover observing times as long as possible in order to allow for detailed time-series analysis, and (4) provide a decent signal-to-noise ratio in order to detect even small ($\lesssim$10\%) variations. Evidently, these criteria are somewhat in conflict for realistic observations: on the hand, criteria 2 and 4 suggest the use of large radio observatories; on the other hand, criterion 3 suggests the exclusive use of a given observatory for long periods, potentially many days. Accordingly, the scheduling of such programs has been notoriously difficult (cf. e.g. \citealt{fuhr08}). 

Our work explores an approach complementary to observations at large observatories. We analyze the performance achievable by using a small, 6-meter class, radio observatory dedicated to monitoring of AGN variability at 230\,GHz. As a realistic example we use the Seoul National University Radio Observatory (SRAO) located in Seoul, Korea, at the geographic position $37^{\circ}27'15''$\,N, $126^{\circ}57'19''$\,E (\citealt{koo03}; Y.-S. Park, \emph{priv. comm.}).

\section{ANALYSIS}

\noindent
Our analysis progresses in three steps. First, we estimate the signal-to-noise ratios (SNR) achievable by an SRAO-type observatory. Second, we select potential target sources based on AGN properties known from the literature. Third, we provide a statistical analysis of various types of AGN variability and the corresponding detection limits. Given the angular resolution of a 6-m radio telescope, we can safely treat all our targets as point sources. Unless stated otherwise, all noise values refer to Gaussian $1\sigma$ levels.

\subsection{Noise Limits}

\begin{table}[t]
\begin{center}
\caption{Parameter values and corresponding noise levels}
\renewcommand\arraystretch{1.2}
\begin{tabular}{c c}
\hline\hline
Parameter & Value  \\
\hline
$A_{\rm e}/A_{0}$ & 0.55 \\
$\Delta t$ & 100\,s \\
$\Delta \nu$ & 1\,GHz \\
$T_{\rm rec}$ & 55\,K \\
$T_{\rm atm}$ & 263\,K \\
$T_{\rm amb}$ & 273\,K \\ % Now Assume T_amb = T_atm + 10K
$\tau_{0}$ & 1.5 \\
$\eta$ & 0.9 \\
$\rho_A$ & 180\,Jy\,K$^{-1}$ \\
$T_{\rm sys}$ & 290\,K \\  %Calculated Again ( += 2K)
$\delta T$ & $0.92$\,mK \\ %Calculated Again ( += 0.01 mK)
$\sigma_N$ & 0.16\,Jy \\ %Same value within two digits accuracy. Thus resulting new figures and numbers would look like reproduction of previous version (In practical calculations , however, I adopted the minor modification; \sigma_{N} = 0.164 Jy --> 0.1656 Jy)
\hline
\label{params}
\end{tabular}
\end{center} 
\end{table}

\noindent
The \emph{signal-to-noise ratio} achieved by a radio observatory can be calculated like

\begin{equation}
\label{snr}
{\rm SNR} = \frac{\rm Signal}{\rm Noise}=\frac{T_{\rm E}}{\delta T}=\frac{T_{\rm E}\sqrt{\Delta \nu \Delta t}}{T_{\rm sys}}
\end{equation}

\noindent
where $T_{\rm E}$ is the equivalent temperature of the source, $\delta T$ is the noise temperature of the telescope, $\Delta \nu$ is the observing bandwidth of the receiver, $\Delta t$ is the integration time, and $T_{\rm sys}$ is the system temperature (cf. \citealt{wilson08}). All temperatures are in units of Kelvin. For a point source, the \emph{equivalent temperature} $T_{\rm E}$ of the source is simply given by \citep{thom04}

\begin{equation}
\label{T_eff}
T_{\rm E}=\frac{A_{\rm e}}{2k}F_{\nu}=\frac{A_{\rm e}}{2800}F_{\nu} \equiv F_{\nu}/\rho_A 
\end{equation}

\noindent
where $k$ is Boltzmann's constant, and $F_{\nu}$ is the flux density of the source in units of Jansky (Jy); the conversion factor $\rho_A$ (in units of Jy\,K$^{-1}$) is usually referred to as \emph{antenna efficiency}. The effective light collecting area $A_{\rm e}$ is given by

\begin{equation}
\label{A_eff}
A_{\rm e} = A_{0}\,e^{-(4\pi \zeta /\lambda)^{2}}
\end{equation}

\noindent
where $A_{0}$ is the physical aperture area, $\zeta$ is the root-mean-squared  difference between actual and ideal antenna surface, and $\lambda$ is the observing wavelength. For the SRAO with an aperture of 6\,m, $A_{\rm e}/A_{0}\approx0.55$ and $\rho_A=180$\,Jy\,K$^{-1}$ at 230\,GHz. The effective aperture is derived from holographic mapping of the antenna surface \citep{koo03}.

The crucial parameter for our analysis is the \emph{system temperature} $T_{\rm sys}$ which can be written as

\begin{equation}
\label{T_sys}
T_{\rm sys}=T_{\rm rec}+T_{\rm atm}\eta\left(1-e^{-\tau_{0}X(\theta)}\right)+T_{\rm amb}\left(1-\eta\right)
\end{equation}

\noindent
where $T_{\rm rec}$ is the receiver noise temperature, $T_{\rm atm}$ is the atmosphere temperature, $\tau_{0}$ is the optical depth of the atmosphere in zenith direction, $X(\theta)$ is the air mass at zenith angle $\theta$, $T_{\rm amb}$ is the ambient temperature, and $\eta$ is the spillover efficiency including rear spillover, scattering, blockage, and ohmic loss efficiency. In the present work we use the approximation $X(\theta)\approx\sec (\theta)$ since we consider only sources with $\theta \lesssim 60^{\circ}$.

In order to arrive at quantitative signal-to-noise ratios, we make the following choices for the values of the various parameters. We assume $\eta=0.9$, in agreement with typical values for modern receivers. For the receiver temperature we assume $T_{\rm rec}=55$\,K; this is five times the quantum noise limit given by $h\nu/k \approx11$\,K and typical for modern receivers (cf. Eq.~5.20 of \citealt{wilson08}; $h$ denotes Planck's constant). In agreement with typical winter conditions at the SRAO site, the remaining temperature terms are assumed to be $T_{\rm atm}=263$\,K and $T_{\rm amb}=273$\,K, with a -- rather conservative -- atmospheric optical depth of $\tau_{0}=1.5$; to take into account an average air mass, we use $\theta=45^{\circ}$. Additionally, we assume a bandwidth of $\Delta\nu=1$\,GHz and select an integration time of $\Delta t=100$\,s. Accordingly, we find $T_{\rm sys}=290$\,K and $\delta T=0.9$\,mK, corresponding to a noise level in units of flux density of $\sigma_N=0.16$\,Jy. All parameters, as well as the resulting noise values, are summarized in Table~\ref{params}.

\subsection{Source Selection}

\begin{deluxetable}{l l r r l r c l}
%\small
\tablecolumns{9}
\tablewidth{0pc}
\tablecaption{Candidate target sources for an SRAO-type radio observatory with SNR$\geq$10}
\tablehead{
\colhead{ID (B1950)} & 
\colhead{Redshift} & 
\colhead{R.A.} & 
\colhead{DEC} & 
\colhead{$F_{\nu}$} & 
\colhead{$\nu_0$} & 
\colhead{SNR} & 
\colhead{Type}
}
\startdata
0059+581$^{\rm e,f}$ & 0.644 & 01h02m45.7s & 58d24m11s & 1.9 & 230 & 12 & LPQ\\
0133+476$^{\rm a}$ &  0.85 &	01h36m58.6s &	48d51m29s & 1.7	& 94 & 11 & HPQ/Blazar/FSRQ\\
0235+164$^{\rm a,b,f}$ & 0.94 & 02h38m38.9s & 16d36m59s & 2.1 & 230 & 13 & FSRQ/BLLac\\
0300+470 & 	0.47 &	03h03m35.2s	& 47d16m16s & 1.7 & 86 & 11 &  BLLac\\
0316+413 & 0.017 & 03h19m48.1s & 41d30m42s & 4.8 & 230 & 32 & Sy2/NLRG\\
0355+508 & 1.52 & 03h59m29.7s & 50d57m05s &4.2 &230 & 27 & LPQ\\ %This source is updated (SNR -= 1)
0415+379 & 0.048 & 04h18m21.3s & 38d01m36s & 4.5 & 230 & 30 & Sy1/BLRG \\
0420$-$014 & 0.91 & 04h23m15.8s & $-$01d20m33s & 2.5 & 231 & 16 & Blazar/BLLac/HPQ  \\
0432+052 & 0.03 & 04h33m11.1s & 05d21m16s & 1.6 & 230 & 10 & LPQ/BLRG \\
0528+134 & 2.07 & 05h30m56.4s & 13d31m55s & 2 & 230 & 13 & LPQ/FSRQ/Blazar \\
0607$-$157$^{\rm c}$ & 0.32 & 06h09m40.9s & $-$15d42m41s & 2.4 & 110 & 14 & FSRQ \\
0727$-$115 & 1.59 & 07h30m19.1s & $-$11d41m13s & 2.3 & 100 & 13 & RLQ  \\
0748+126 & 0.88 & 07h50m52.0s & 12d31m05s & 2.3 & 230 & 14 & LPQ/FSRQ  \\ %This source is updated (SNR -= 1)
0827+243 & 0.94 & 08h30m52.1s & 24d11m00s & 1.6 & 96 & 10 & FSRQ/LPQ/Blazar \\
0851+202 & 0.3 & 08h54m48.9s & 20d06m31s & 4 & 230 & 26 & BLLac \\
0923+392 & 0.69 & 09h27m03.0s & 39d02m21s & 2.6 & 230 & 17 & Sy1/LPQ  \\
1030+415 & 1.11 & 10h33m03.7s & 41d16m06s & 1.5 & 94 & 10 & HPQ/FSRQ \\
1055+018 & 0.89 & 10h58m29.6s & 01d33m59s & 2.5 & 230 & 15 & HPQ/BLLac \\
1253$-$055 & 0.53 & 12h56m11.1s & $-$05d47m22s & 9.5 & 230 & 58 & BLLac/HPQ/FSRQ\\ 
1334$-$127 & 0.53 & 13h37m39.8s & $-$12h57m25s & 3.8 & 230 & 22 & HPQ/BLLac\\
1510$-$089 & 0.36 & 15h12m50.5s & $-$09d06m00s & 1.6 & 230 & 10 & Sy1/HQP\\
1606+106 & 1.22 & 16h08m46.2s & 10d29m08s & 1.8 & 94 & 11 & FSRQ/LPQ/Blazar\\
1633+382 & 1.81 & 16h35m15.5s & 38d08m04s & 1.8 & 230 & 12 & FSRQ/LPQ/Blazar\\
1641+399 & 0.592 & 16h42m58.8s & 39d48m37s & 2.8 & 230 & 18 & HPQ/FSRQ\\
1642+690$^{\rm a,b,f}$ & 0.75 & 16h42m07.8s & 68d56m40s & 2.3 & 230 & 14 & HPQ/FSRQ\\
1730$-$130 & 0.902 & 17h23m02.7s & $-$13d04m50s & 2.4 & 230 & 14 & BLLac/FSRQ/LPQ\\
1749+096 & 0.322 & 17h51m32.8s & 09d39m01s & 2.7 & 230 & 17 & HPQ/BLLac\\
2013+370 & --- & 20h15m28.7s & 37d11m00s & 2.2 & 230 & 14 & BLLac\\
J2044$-$1043 & 0.034 & 20h44m09.7s & $-$10d43m25s & 2.1 & 190 & 12 & Sy1.2\\ %This source is updated (SNR -= 1)
2145+067 & 0.99 & 21h48m05.4s & 06d57m39s & 1.6 & 230 & 10 & LPQ/FSRQ\\
2200+420$^{\rm d}$ & 0.068 & 22h02m43.3s & 42d16m40s & 1.8 & 230 & 12 & BLLac \\
2216$-$038 & 0.901 & 22h18m52.0s & $-$03h35m37s &  1.6 & 262 & 10 & FSRQ/LPQ\\
2223$-$052 & 1.4 & 22h25m47.2s & $-$04d57m01s & 2.4 & 230 & 14 & HPQ/BLLac\\
\enddata
\label{targets}
\tablenotetext{\hspace{-3.25mm}}{{\sc Notes:} $F_{\nu}$ is the source flux at 230GHz in units of Jy, $\nu_0$ is the observation frequency of data taken from the NED, in units of GHz. Entries ``---'' denote ``not available''. For some objects IDV has been reported by the studies referred to below.\vspace{1mm}}
\tablenotetext{\hspace{-3.25mm}}{{\sc References:} a --- \citealt{qui92}; b --- \citealt{qui2000}; c --- \citealt{ked01}; d --- \citealt{kra03}; e --- \citealt{lov03}; f --- \citealt{oj04}}
\end{deluxetable}

\begin{figure*}[!t]
\centering \epsfxsize=14cm
\epsfbox{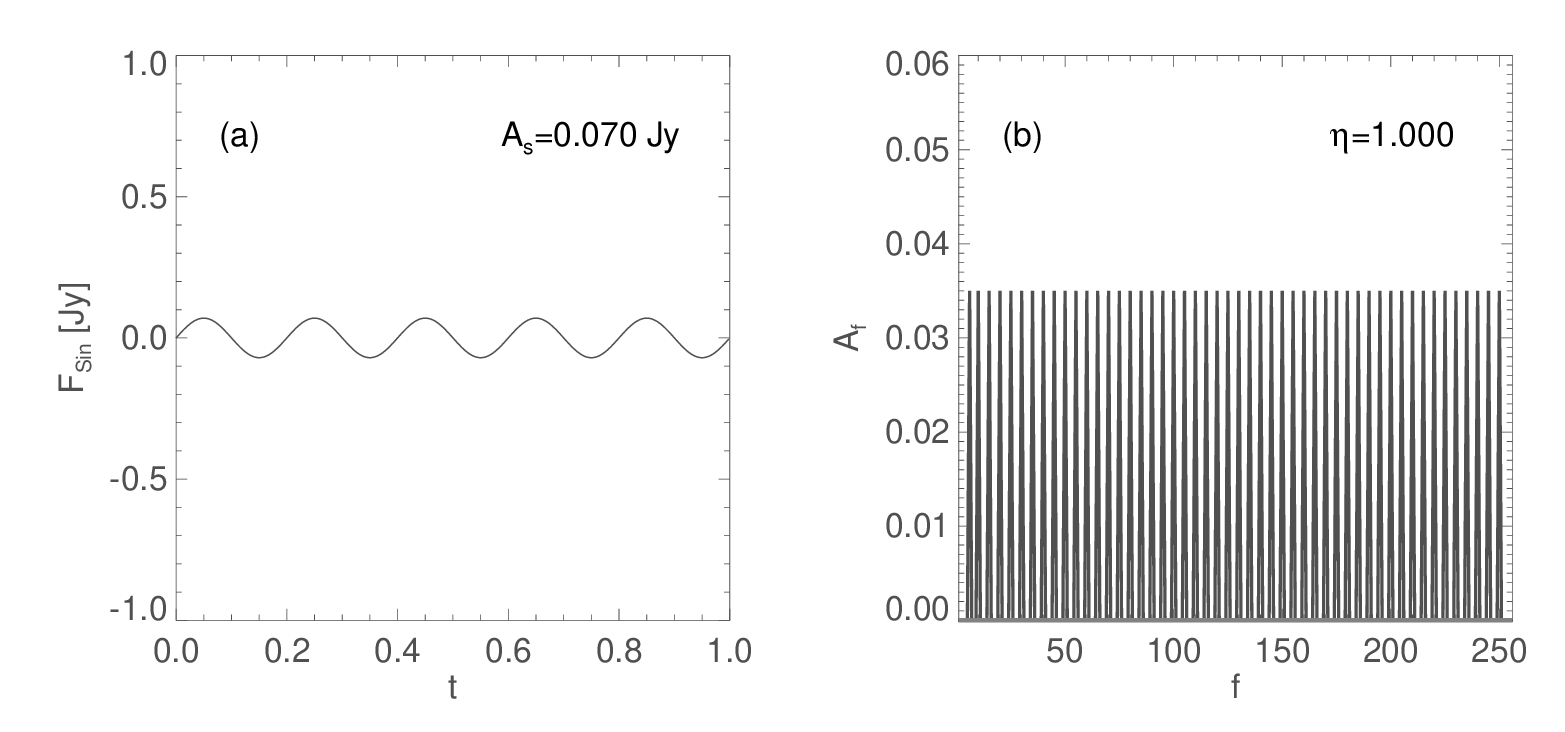} \\
\centering \epsfxsize=14cm
\epsfbox{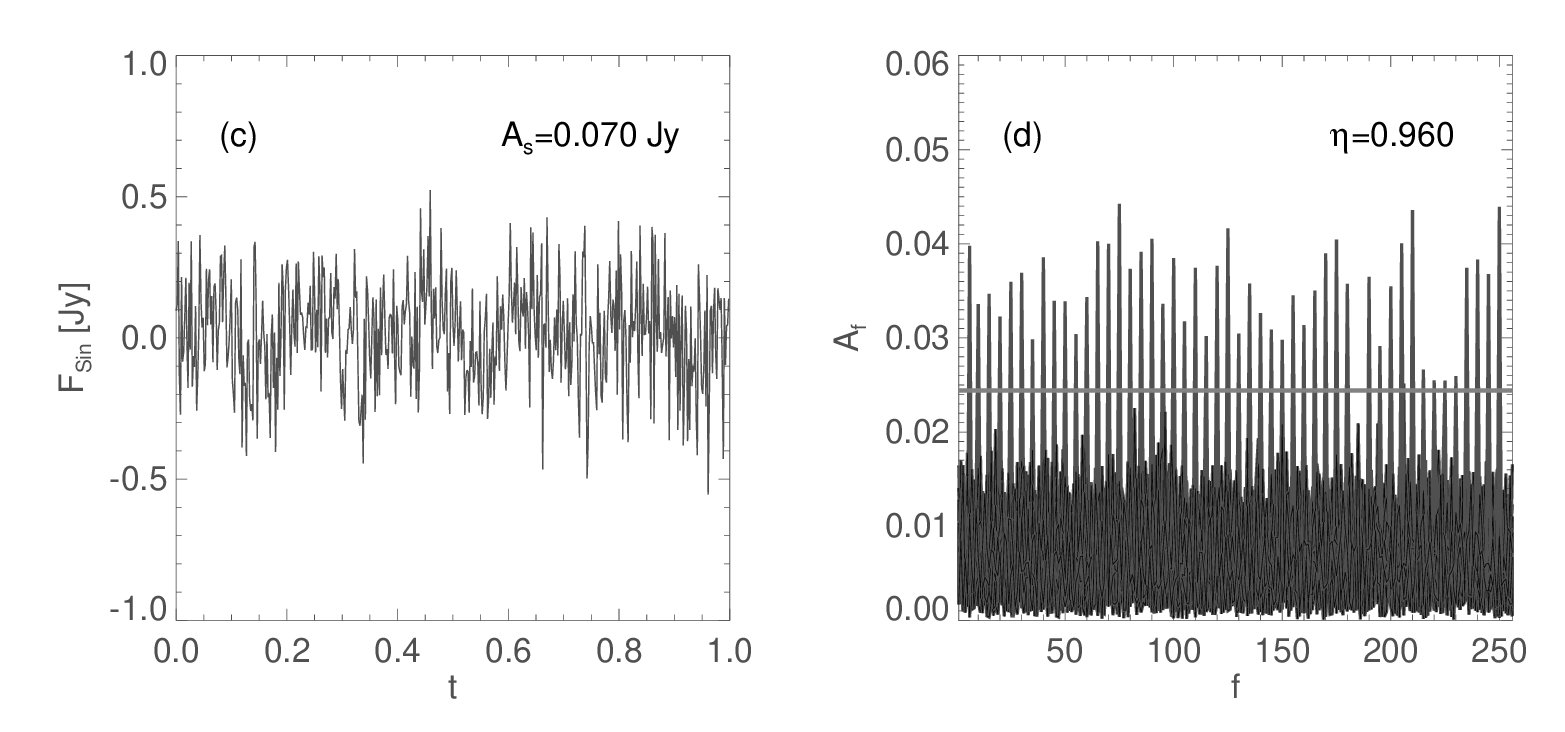} \\
\centering \epsfxsize=14cm
\epsfbox{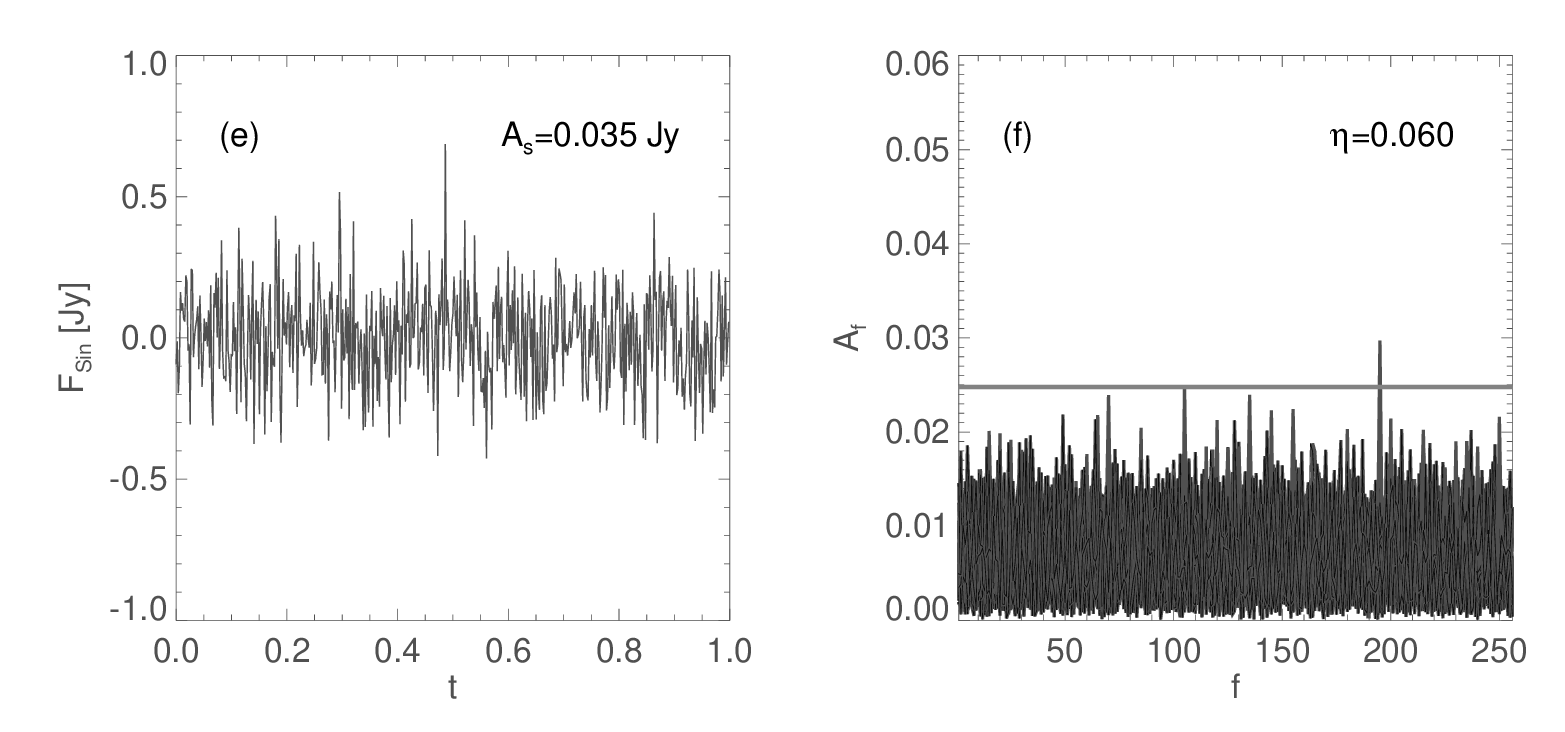}
\caption{Illustration of the analysis scheme for {\sc periodic variability}. {\bf a.} Idealized, noise-free variable lightcurve with amplitude $A_s=0.07$\,Jy and variability frequency $f=5$. {\bf b.} Overlay of 50 Fourier transforms of noise-free lightcurves with $A_s=0.07$\,Jy and $f=5, 10, ...,250$; as the signals are free of noise, the detection efficiency $\eta$ is unity. {\bf c.} Like {\bf a} but with Gaussian noise with amplitude $\sigma_N=0.16$\,Jy. {\bf d.} Like {\bf b} but for lightcurves with noise $\sigma_N$. The horizontal grey line marks the $5\sigma$ detection threshold. {\bf e.} Like {\bf c} but with variability amplitude reduced to $A_s=0.035$\,Jy. {\bf f.} Like {\bf d} but for lightcurves with $A_s=0.035$\,Jy. The substantial decrease in detection efficiency between {\bf d} and {\bf f} is obvious.}
\label{periodic} %Fig.1; Drawn again by using almost same (but ~1% different) noise flux value. 
\end{figure*}

\noindent
In order to select a set of candidate target sources, we first extracted a list of 407 radio-bright AGN from the NASA/IPAC Extragalactic Database (NED). Unfortunately, not for all sources data taken at 230\,GHz are available. In those cases, we obtained reference fluxes $F_{\nu_{0}}$, where $\nu_{0}$ is a frequency as close to 230\,GHz as possible. We made the usual assumption that the radio spectra of our target AGN follow a power law

\begin{equation}
\label{powerlaw}
F_{\nu}=F_{\nu_{0}}\left(\frac{\nu}{\nu_{0}}\right)^{-\alpha}
\end{equation}

\noindent
where $F_{\nu}$ is the source flux at frequency $\nu$, $F_{\nu_{0}}$ is the source flux at frequency $\nu_{0}$, and $\alpha$ is the spectral index. We assumed a typical spectral index $\alpha=0.6$ to estimate the source fluxes at 230\,GHz.

We aim at target sources with SNR sufficient to detect rapid variability. Accordingly, our final sample is composed of sources with (1) signal-to-noise ratios larger or equal than ten, and (2) transit zenith angles less than $60^{\circ}$. For each source we computed an individual $\sigma_N$ using the zenith angle at transit for reference and from this the corresponding signal-to-noise ratio. For the location of the SRAO, this left us with 33 sources. The source with the highest SNR turned out to be 1253$-$055 ($F_{\rm 230\,GHz}\approx9.5$\,Jy, SNR$\approx$58), the source with the lowest SNR was 1030+415 ($F_{\rm 230\,GHz}\approx1.5$Jy, SNR$\approx$10). %%% Name of the source with loweest SNR is changed (please check it with previous version). I don't know how I missed this in the last checking but I think lowest SNR case is 1030+415 considering its lowest flux(1.5Jy) and calculated SNR(10, the marginal value)
\noindent We summarize our sample in Table~\ref{targets}; we note that for five of our target AGN intra-day variability has already been reported in the literature. 

We employed various checks to ensure the reliability of our sample. Modifying the values for various critical parameters, especially $\alpha$, changes the number of selected targets by less than 20\%. Therefore we conclude that our sample is fairly insensitive to the choice of parameters.

\subsection{Detection Limits of Variability}

\noindent
Our study also aims at a quantitative description of the variability that might be detected under the conditions outlined above. Accordingly, we analyze simulated, though realistic, observations statistically by means of Monte Carlo simulations. In the following, we always assume a single long -- $\mathcal{T}$=$14.2$\,h -- observation of an arbitrary target. The observation time is divided into 512 bins of 100\,s each. Our calculations address two types of potential variability: a periodic flux modulation (periodic variability) and a single outburst of activity (flaring variability). Calculations are performed in {\tt IDL}.\footnote{Interactive Data Language, ITT Exelis Inc., McLean (Virginia)}

\subsubsection{Periodic Variability}

\noindent
An artificial lightcurve corresponding to a noisy signal with periodic variability and zero average, $F_{\rm sin}$, is created by

\begin{equation}
\label{F_var_sin}
F_{\rm sin} = A_{\rm s}\,\sin(2\pi f t)+F_{\rm noise}
\end{equation}

\noindent
where $A_{\rm s}$ is the amplitude of the modulation, $f$ is the frequency of the periodic modulation, $0\le t\le1$ is the time in units of the total observing time $\mathcal{T}$, and $F_{\rm noise}$ is Gaussian random noise with amplitude $\sigma_N$.

\begin{figure}[!t]
\centering \epsfxsize=8.25cm
\epsfbox{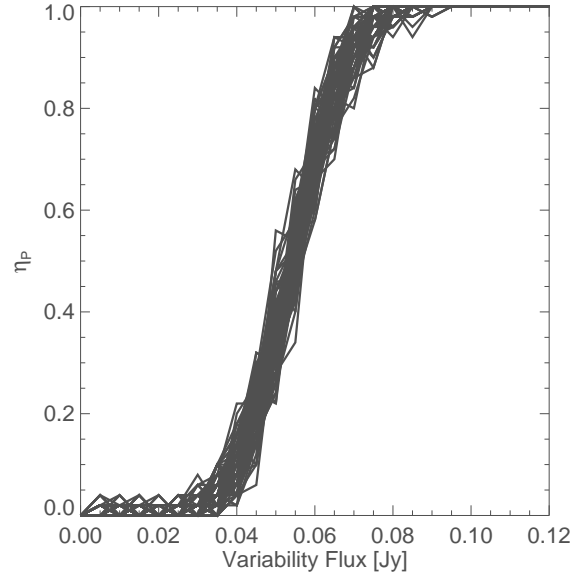}
\caption{Detection efficiency $\eta_P$ of {\sc periodic variability} as function of amplitude of flux variation. This diagram is an overplot of 100 analysis runs (grey lines); for each flux amplitude, 100 values of $\eta_P$ are computed, thus showing the confidence ranges.}
\label{eff_res} %Fig.2 ; Drawn again.
\end{figure}

We analyze any simulated lightcurve by taking the absolute value of its Fourier transform. We classify a flux variability as detected if the Fourier transform has a peak exceeding its average by five times its standard deviation.\footnote{This straightforward approach is possible because our artificial data are evenly sampled. Realistic flux observations usually require the use of periodograms for which the false alarm probability follows an exponential distribution \citep{scargle82}.} In order to derive detection thresholds, we examine 50 different amplitudes $A_{\rm s} = 0.005, 0.01, ..., 0.25$\,Jy. For each amplitude, we probe 50 different frequencies $f = 5, 10, ..., 250$.\footnote{As we normalize the total observing time to the range $[0,1]$, the frequencies are simply unit-free integers. The Nyquist frequency is half the number of data points, i.e. 256 in our case.}  For a given amplitude $A_s$, we define a \emph{detection efficiency}

\begin{equation}
\label{eff_P}
\eta_{\rm P}(A_{\rm s})=\frac{N_{\rm d,P}}{50}
\end{equation}

\noindent
where $N_{\rm d,P}$ is the number of detections which is 50 at best (for 50 frequencies of periodic variability). In order to obtain statistically robust results, we test each combination of $f$ and $A_s$ 100 times, meaning we analyze 250\,000 simulated lightcurves in total. We illustrate our analysis procedure in Fig.~\ref{periodic} and present the detection efficiency as function of amplitude in Fig.~\ref{eff_res}.

\subsubsection{Flaring Variability}

\begin{figure}[!t]
\centering
\epsfxsize=8.25cm
\epsfbox{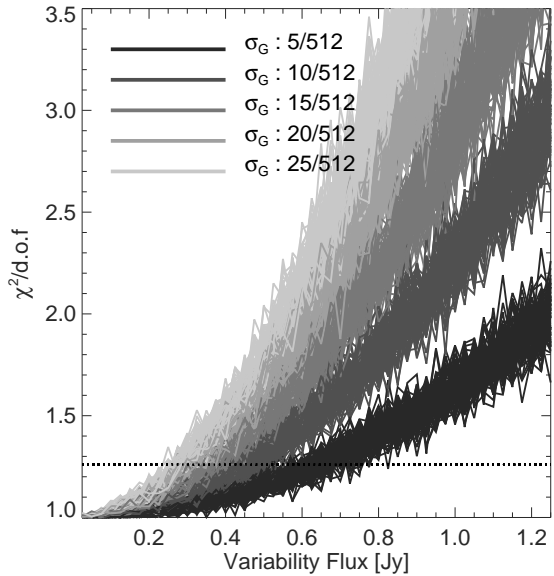}
\caption{Reduced $\chi^{2}$ test probing the detection of {\sc flaring variability}. The reduced $\chi^2$ is shown here as function of variability flux $A_G$, for five different flare time scales $\sigma_G$. For each $\sigma_G$, 100 curves -- one for each realization -- are plotted on top of each other. The horizontal dotted line marks the $\chi^{2}/{\rm d.o.f.}=1.26$ limit above which significant variability is detected.}
\label{chi_res}
\end{figure} % Fig.3 ; Drawn again.

\begin{figure*}[!t]
\centering \epsfxsize=14cm
\epsfbox{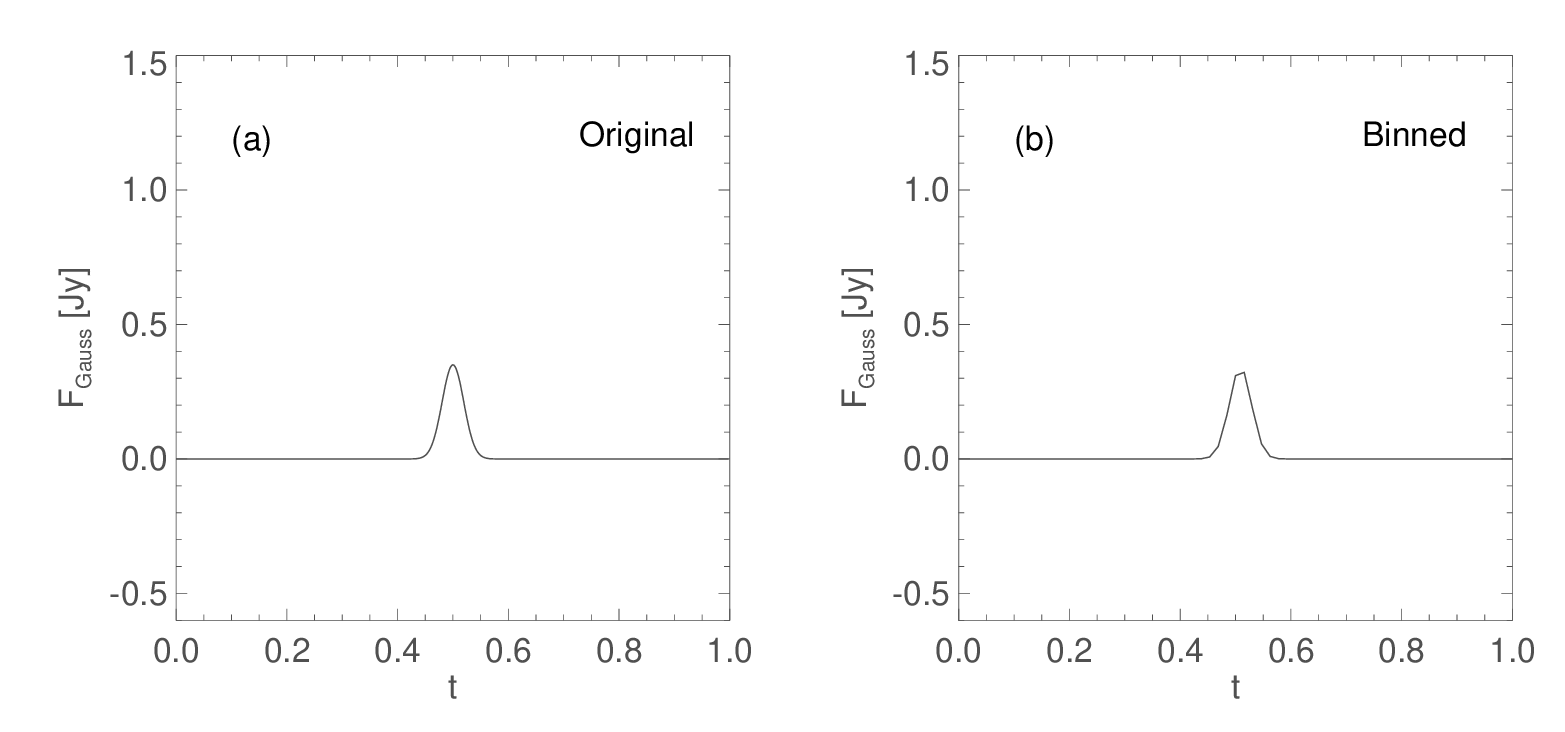} \\
\centering \epsfxsize=14cm
\epsfbox{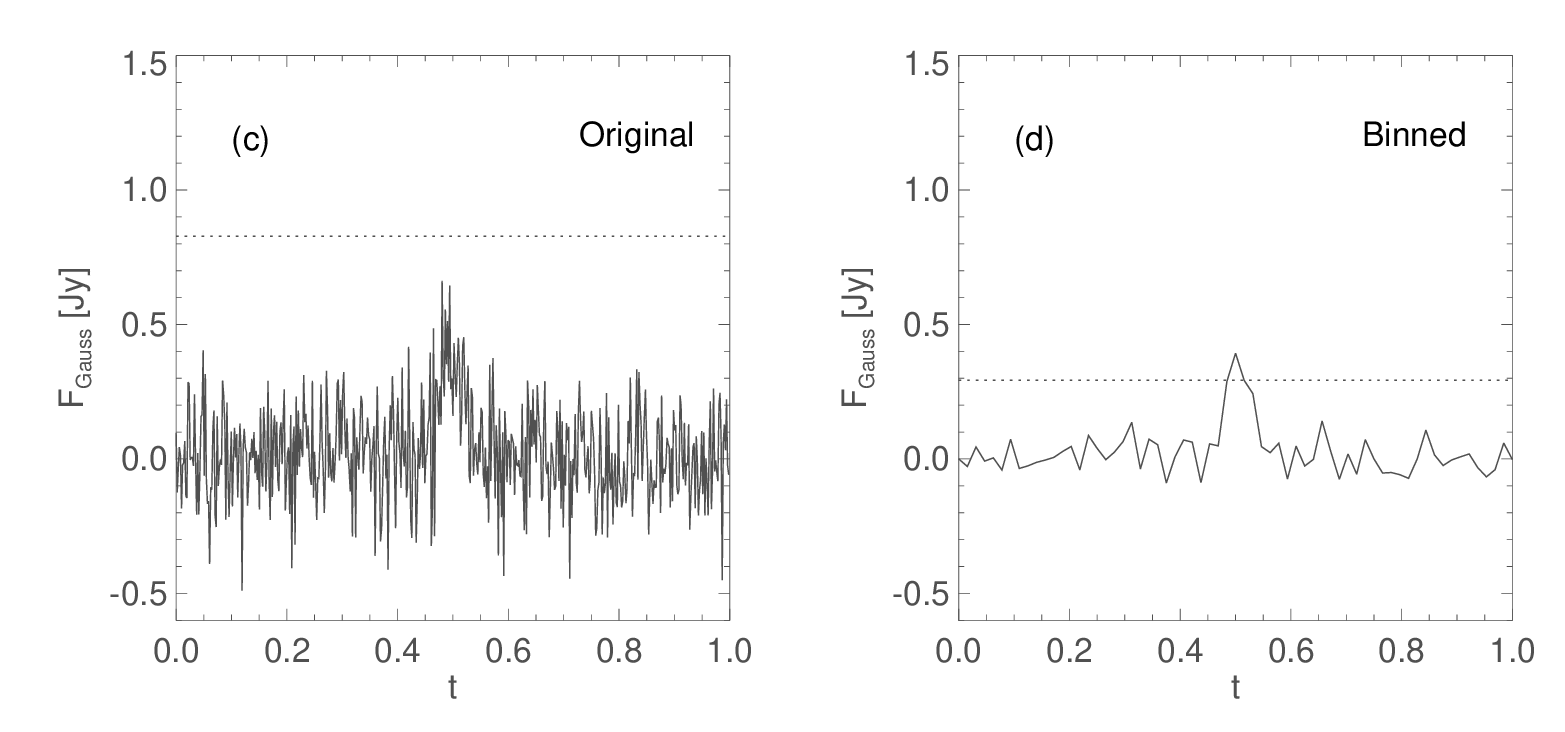} \\
\centering \epsfxsize=14cm
\epsfbox{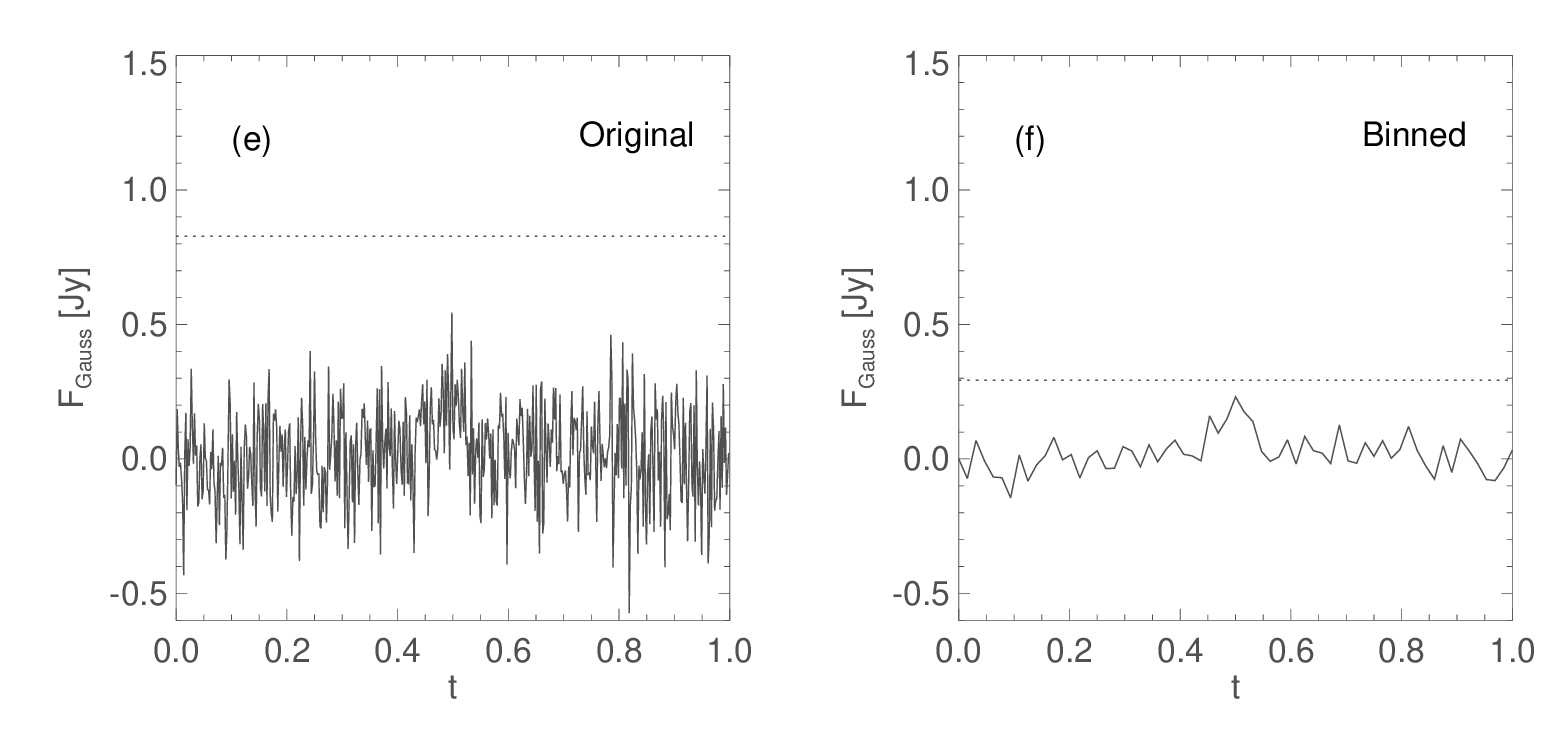}
\caption{Illustration of the {\sc multi-scale analysis} scheme for {\sc flaring variability}. {\bf a.} Idealized, noise-free lightcurve with a Gaussian flux peak with amplitude $A_G=0.35$\,Jy and width $\sigma_G=20/512$. {\bf b.} The lightcurve {\bf a} with $\mathcal{N}$=$8$ adjacent data points being binned into one. {\bf c.} Lightcurve {\bf a} with Gaussian noise with amplitude $\sigma_N=0.16$\,Jy added. {\bf d.} Binned version of lightcurve {\bf c}. {\bf e.} Like lightcurve {\bf c} but with flux peak amplitude reduced to $A_G=0.17$\,Jy. {\bf f.} Binned version of {\bf e}. In all diagrams the horizontal dotted lines mark the respective $5\sigma$ detection thresholds. The impact of binning and flux peak amplitude is obvious.}
\label{gauss}
\end{figure*} %Fig.4 ; Drawn again.

\begin{figure*}[!t]
\centering
\epsfxsize=17.5cm
\epsfbox{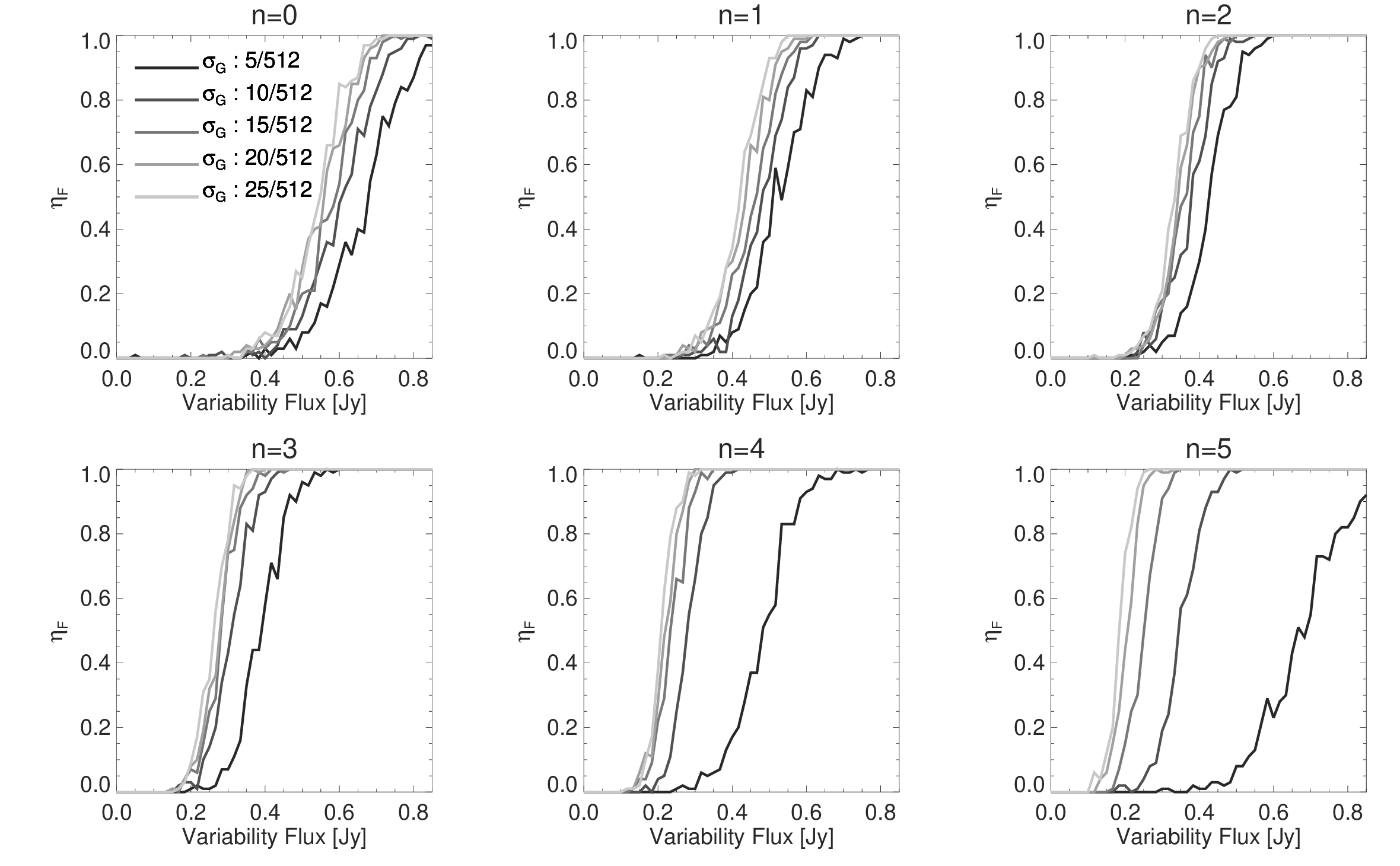}
\caption{Detection efficiency $\eta_F$ for {\sc flaring variability} as function of flux peak amplitude (variability flux) and binning level $n$ derived from {\sc multi-scale analysis}. Each diagram shows five curves corresponding to the five flare time scales indicated in the top left panel.}
\label{res_all}
\end{figure*} % Fig. 5 ; Drawn again.

\noindent
In order to create an artificial lightcurve with episodic, flaring variability, we parametrize the flare as a singular flux peak with Gaussian shape. Accordingly, our simulated time series is

\begin{equation}
\label{F_var_gauss}
F_{\rm Gauss} = A_{\rm G}\,\exp\left[-\frac{(t-0.5)^{2}}{2\sigma_{\rm G}^{2}}\right]+F_{\rm noise}
\end{equation}

\noindent
where $A_G$ is the flare amplitude, $t$ is the time in units of total observing time, $\sigma_G^2$ is the temporal variance of the flux peak (variability time scale), and $F_{\rm noise}$ is Gaussian random noise with amplitude $\sigma_N$. The flare is centered at $t=0.5$. This parametrization takes into account explicitly that the detectability of a flux peak depends on its amplitude as well as its time scale. We examine 50 different flare amplitudes $A_{\rm G}$ and five flare time scales $\sigma_{\rm G}=5/512, 10/512, ..., 25/512$, meaning 250 parameter combinations in total.

\paragraph{Reduced $\chi^2$ Test.}

Else than for periodic variability, there is no obvious recipe for the detection of flaring flux modulations. A simple approach is provided by a $\chi^2$ test that uses a signal constant in time (absence of variability) as null hypothesis. Taking into accounts the number of degrees of freedom leads to a \emph{reduced} $\chi^2$ test

\begin{equation}
\label{chi_cal}
\chi^{2}/{\rm d.o.f.} = \frac{1}{N-1}\sum_{i=1}^{N} \frac{F^2_{\rm Gauss,\it i}}{\sigma_{\rm N}^{2}}
\end{equation}

\noindent
where $F_{\rm Gauss,\it i}$ is the $i$-th value of the lightcurve defined by Eq.~\ref{F_var_gauss}, d.o.f. denotes the number of degrees of freedom, and $N$ is the number of data points. In case of absence of variability, $\chi^{2}/{\rm d.o.f.}\approx1$ by construction. We calculate 100 realizations of $\chi^{2}/{\rm d.o.f.}$ for each combination of $A_G$ and $\sigma_G$, meaning 25\,000 calculations in total. We examine a range of flare amplitudes $A_{\rm G}=0.025, 0.05, ..., 1.25\,$Jy. In Fig.~\ref{chi_res} we show $\chi^{2}/{\rm d.o.f.}$ as function of $A_G$ for the five flare time scales examined. We consider the null hypothesis ``the lightcurve is not variable'' as rejected if $\chi^{2}/{\rm d.o.f.}\geq1.26$, corresponding to a Gaussian significance of $5\sigma$ for 511 degrees of freedom. With increasing flare duration, i.e. increasing number of data points affected, the minimum flare amplitude necessary for detection decreases.

\paragraph{Multi-Scale Analysis.}

A more sophisticated but nevertheless straightforward approach to detecting flaring variability is provided by \emph{multi-scale analysis} (e.g. \citealt{starck2006}). The underlying scheme is the following:

\begin{enumerate}

\item Take a simulated lightcurve and search for a signal in excess of five times the noise level of the data.

\item Bin the lightcurve such that $\mathcal{N}$ adjacent data points are combined into one averaged value.

\item Repeat step 1.

\end{enumerate}

\noindent
We apply this scheme to our data in an iterative fashion. In each binning level $n=0, 1, 2, 3, 4, 5$ we combine two adjacent data points such that $\mathcal{N}$=$2^n$ data points of the original lightcurve have been combined after step $n$; the case $n=0$ simply corresponds to the original time series. Binning reduces the noise in a lightcurve from $\sigma_N$ to $\sigma_N/\sqrt{\mathcal{N}}$. Using our usual $5\sigma$ criterion for counting a detection as significant, we are able to detect flux peaks with amplitudes $5\sigma_N/\sqrt{\mathcal{N}}$ -- provided the flux peak actually covers approximately $\mathcal{N}$ data points in the original lightcurve. We illustrate our analysis scheme in Fig.~\ref{gauss}.

We analyze 100 realizations for each combination of flare amplitude $A_G$, flare time scale $\sigma_G$, and binning level $n$, meaning 150\,000 lightcurves in total. We examine a range of flare amplitudes $A_{\rm G}=0.017, 0.034, ..., 0.85\,$Jy. We define a detection efficiency

\begin{equation}
\label{eff_F}
\eta_{\rm F}(A_{\rm G},\sigma_{\rm G},n)=N_{\rm d,F}/100
\end{equation}

\noindent
where $N_{\rm d,F}$ is the number of times variability is detected out of 100 trials for a given set of $A_G$, $\sigma_G$, and $n$. The resulting detection efficiencies are shown in Fig.~\ref{res_all}.

\section{DISCUSSION}

\noindent
Our analysis probes the capabilities of a 6-meter class radio telescope with respect to continuum flux observations of AGN intra-day variability at high radio frequencies around 230\,GHz. Assuming typical weather condition for the location of the SRAO and assuming the validity of the usual noise laws of radio astronomy, we conclude that an SRAO-type observatory is affected by a statistical ($1\sigma$) noise limit of $\sigma_N\approx0.16$\,Jy. Demanding a signal-to-noise ratio $\geq10$ for AGN flux monitoring, we are able to identify 33 sources observable from the location of the SRAO (Table~\ref{targets}). Thus we may conclude that an SRAO-type observatory is indeed able to make valuable contributions to monitoring studies of AGN.

We note that our calculations throughout Sect.~2 assume somewhat idealized observations leading to densely sampled lightcurves without interruptions. In realistic observations, the need for occasional calibrations of flux and amplitude, the observing schemes, and technical boundary conditions will result in more complicated datasets with details depending on the characteristics of the observatory actually employed. Our results depend only weakly on our choice for the total observing time $\mathcal{T}$=$14.2$\,h: shorter observing times reduce the maximum time scale of observable flux variations, but not the sensitivity (which is calculated for 100\,s of integration time).

Unsurprisingly, the limit for the detection of flux variations depends on the morphology of the variability. We analyze two rather extreme examples, namely periodic and flaring flux modulations. The detection efficiency (Figs.~\ref{periodic}, \ref{eff_res}) of \emph{periodic} variability -- which takes into account many different frequencies of modulation up to almost the Nyquist limit -- approaches unity when the amplitude of flux modulations exceeds approximately 0.08\,Jy. The detection limits for \emph{flaring} variability show a strong interplay between the amplitude of a flux peak and its duration: shorter flares require larger amplitudes in order to be detected. A simple $\chi^2$ test (Fig.~\ref{chi_res}) indicates that the minimum amplitudes necessary are approximately located between 0.3\,Jy and 0.8\,Jy for characteristic flare time scales $\sigma_G$ between $25/512$ (in units of total observing time, corresponding to $\approx$42\,min) and $5/512$ ($\approx$8\,min). Using a more sophisticated multi-scale analysis and demanding detection efficiencies close to unity (Figs.~\ref{gauss}, \ref{res_all}) reduces these limits to the range from $\approx$0.3\,Jy to $\approx$0.5\,Jy. In realistic AGN lightcurves, variability follows red noise power laws (e.g. \citealt{trippe2011,park2012}) with stronger flux modulations occurring on longer time scales; red noise variability may be regarded as a case intermediate between periodic and flaring variability. Notably, we may conclude from our analyses that we are able to detect significant variability with amplitudes $\lesssim2\sigma_N\approx0.3$\,Jy, i.e. on levels $\lesssim10$\% of the average fluxes of our targets.

The physical conditions within a target AGN can be probed via analysis of timescales and amplitudes of flux variations. Using realistic astrophysical parameter values, the observed brightness temperature can be expressed \citep{wag95} like

\begin{equation}
\label{T_br}
T_{\rm b}=4.5\times10^{10}\,A_{\rm var}\left[\frac{\lambda\,d}{t_{\rm var} (1+z)}\right]^{2} ~ {\rm K}
\end{equation}

\noindent
where $A_{\rm var}$ is the amplitude of a flux modulation (in units of Jy), $\lambda$ is the wavelength (in cm), $d$ is the distance to the source (in Mpc), $t_{\rm var}$ is the variability timescale (in days), and $z$ is the redshift of the source. For a source at $z=0.3$ with $A_{\rm var}=0.3$\,Jy and $t_{\rm var}=40$\,min, $T_b\approx10^{17}$\,K for observations at 230\,GHz. Even for variability timescales on the order of one day, $T_b\approx10^{14}$\,K. Accordingly, any detection of intra-day variability with an SRAO-type radio observatory corresponds to the observations of sources with observed brightness temperatures exceeding the Inverse-Compton limit. Using the common assumption that such an excess in observed brightness temperature is due to Doppler boosting in AGN outflows (e.g. \citealt{fuhr08}) one finds an expression for the Doppler factor $\delta$ like

\begin{equation}
\label{eq_doppler}
\delta = (1+z) \left(\frac{T_b}{10^{12}\,{\rm K}}\right)^{1/(3-\alpha)}
\end{equation}

\noindent
where $z$ is the redshift of the source and $\alpha$ is the spectral index (cf. Eq.~\ref{powerlaw}). For observed $T_b\approx10^{14...17}$\,K, $\delta\approx8...150$ -- meaning that we are, a priori, able to probe a wide range of plasma-physical and kinematic conditions within AGN.

High-frequency radio monitoring of radio-bright compact AGN is an important complement to simultaneous observations at other wavelengths, both for single-dish flux monitoring (e.g. \citealt{fuhr08}) as well as radio-interferometric mapping (e.g. \citealt{lee2008}). An important example are measurements of time lags between lightcurves obtained at different frequencies, providing the dispersion measure -- the line-of-sight integral of the particle density -- and thus information on the matter content of the emission region (e.g. \citealt{wilson08}). Accordingly, the combination  of small, 6-m class observatories working at high frequencies with larger observatories working at lower frequencies -- like the Korean VLBI Network (KVN) which operates at frequencies of 22, 43, 86, and 129\,GHz \citep{kim2011,lee2011} -- is likely to provide new insights into AGN physics.

\section{CONCLUSIONS}

\noindent
In this article, we study the feasibility of high-frequency radio observations of AGN intra-day variability. We assume the use of a small 6-meter class observatory for continuum flux monitoring at 230\,GHz, using the SNU Radio Astronomical Observatory (SRAO) as a realistic example. Our work arrives at the following principal conclusions:

\begin{enumerate}

\item  Assuming an integration time of 100\,s per flux measurement, a spectral bandwidth of 1\,GHz, and typical SRAO winter conditions, the statistical $1\sigma$ noise limit is $\sigma_N\approx0.16$\,Jy.

\item  Demanding a signal-to-noise ratio $\geq10$, we compile a list of 33 radio-bright compact AGN with fluxes $F_{\nu}\gtrsim1.5$\,Jy that are observable at the geographic location of the SRAO.

\item  Based on results from exhaustive Monte-Carlo simulations, we conclude that it is possible to detect significant (referring to a $5\sigma$ detection threshold) variability with amplitudes $\lesssim0.3$\,Jy, roughly 10\% of the typical fluxes of our candidate target sources. Details depend on the morphology of variability -- periodic, episodic/flaring, or intermediate.

\item  For nearby AGN (using here $z=0.3$ for reference), observable flux modulations correspond to observed brightness temperatures $T_b\approx10^{14...17}$\,K, meaning Doppler factors of $\delta\approx8...150$. This indicates that observations of the type we study are able to probe a wide range of plasma-physical and kinematic conditions within AGN.

\end{enumerate}

Overall, we are able to conclude that high-frequency radio observations of AGN intra-day variability with dedicated small observatories can provide valuable new insights into AGN physics. This approach is especially powerful when combining those high-frequency radio observations with data from other facilities like the KVN.

\acknowledgments{\noindent\small We are grateful to {\sc Yong-Sun Park} for enlightening discussion and to {\sc Junghwan Oh} (both at SNU) for valuable technical support. We acknowledge financial support from the Korean Astronomy and Space Science Institute (KASI) via Research Cooperation Grant 2012-1-600-90. Our work has made use of the NASA/IPAC Extragalactic Database (NED) which is operated by the Jet Propulsion Laboratory, California Institute of Technology, under contract with the National Aeronautics and Space Administration of the USA. Last but not least, we are greatful to an anonymous reviewer whose careful report helped to improve this paper.}

{}

\end{document}